\documentclass[fleqn,usenatbib]{mnras}

\usepackage{newtxtext,newtxmath}

\usepackage[T1]{fontenc}

\DeclareRobustCommand{\VAN}[3]{#2}
\let\VANthebibliography\thebibliography
\def\thebibliography{\DeclareRobustCommand{\VAN}[3]{##3}\VANthebibliography}

\usepackage{graphicx}	
\usepackage{amsmath}	
\usepackage{aas_macros}

\title[On The Distribution of Bayesian Evidences]{On The Distribution of Bayesian Evidences}

\author[R. E. Keeley et al.]{
Ryan E. Keeley,$^{1,2}$\thanks{E-mail: rkeeley@ucmerced.edu}
Arman Shafieloo,$^{2,3}$\thanks{E-mail: shafieloo@kasi.re.kr}
\\
$^{1}$Department of Physics, University of California Merced, 5200 North Lake Road, Merced, CA 95343, USA\\
$^{2}$Korea Astronomy and Space Science Institute (KASI),\\776 Daedeok-daero, Yuseong-gu, Daejeon 34055, Korea\\
$^{3}$KASI Campus, University of Science and Technology,\\217 Gajeong-ro, Yuseong-gu, Daejeon 34113, Korea
}

\date{Accepted XXX. Received YYY; in original form ZZZ}

\pubyear{2022}

\begin{document}
\label{firstpage}
\pagerange{\pageref{firstpage}--\pageref{lastpage}}
\maketitle

\begin{abstract}
We look at the distribution of the Bayesian evidence for mock realizations of supernova and baryon acoustic oscillation data.  The ratios of Bayesian evidences of different models are often used to perform model selection.  The significance of these Bayes factors are then interpreted using scales such as the Jeffreys or Kass \& Raftery scale. First, we demonstrate how to use the evidence itself to validate the model, that is to say how well a model fits the data, regardless of how well other models perform.  The basic idea is that if, for some real dataset a model's evidence lies outside the distribution of evidences that result when the same fiducial model that generates the datasets is used for the analysis, then the model in question is robustly ruled out. Further, we show how to assess the significance of a hypothetically computed Bayes factor.  We show that the range of the distribution of Bayes factors can greatly depend on the models in question and also the number of data points in the dataset. Thus, we have demonstrated that the significance of Bayes factors needs to be calculated for each unique dataset.  
\end{abstract}

\begin{keywords}
dark energy -- cosmological parameters -- methods: statistical
\end{keywords}



\section{\label{sec:Intro}Introduction}
Modern cosmology, in its current ``precision era'', requires a thorough understanding of the employed statistical methodology in order to make robust inferences from the data.  This is especially important with regards to understanding the plethora of statistical ``tensions'' that have cropped up in the past years, which have the potential to challenge the current concordance model of cosmology $\Lambda$CDM (cosmological constant dark energy; cold dark matter).  The most notorious tension is the ``$H_0$ tension'' which is the disagreement between the local expansion rate inferred by Planck's measurement of the cosmic microwave background (CMB)~\cite{Planck18} and that value directly measured by observations of Cepheids by the SH0ES collaboration~\cite{riess-etal19}.  These tensions are often investigated in terms of a model selection question where the $\Lambda$CDM model is compared to some new model~\cite{2021A&A...647L...5J,2021A&A...649A..88T}.  The comparison is typically performed by computing the Bayes factor $B_{ij} = \log{Z_i/Z_j}$ where $Z$ is evidence, $P(D|M_{i , j})$, the probability of the data, $D$, given model $M_{i , j}$. If $B_{ij} > 0$ the model $i$ is favored over model $j$ and vice versa. This quantity is useful since, using Bayes' theorem $P(M_i|D) = P(D|M_i)P(M_i)/P(D)$, and using equal prior probabilities for the models, implies that $B_{ij} = \log{P(M_i|D) / P(M_j|D)}$. Thus the Bayes factor corresponds to the relative probability of one model over another.

Seminal works which have laid the ground work for the adoption of Bayesian statistics in cosmology include~\citet{2007MNRAS.378...72T,2007MNRAS.377L..74L,2008ConPh..49...71T}. In particular, \citet{2007MNRAS.378...72T} introduces the Savage-Dickey ratio which is an efficient way to calculate the Bayes factor for the case where the two models in question are nested.

There are a few drawbacks to this inference procedure.  The first problem is that this procedure is inherently a comparison; it can only determine which model is better or worse, not whether either model is a good fit to the data in an absolute sense. In other words, it is useful only for model selection, not for model validation.  This is an important point to keep in mind since there is no reason to believe that the true model of the Universe, whatever it may be, could be guessed at and so included as a discrete choice in this model selection procedure.  

The second drawback is a question of significance.  The mathematics of the procedure returns a number.  It is a further step of human interpretation to decide what that number means in terms of whether we should believe one model is true and the other is false.  This interpretation is done with one of the two commonly used scales, the Jeffreys scale~\cite{jeffreysscale} and the Kass \& Raftery scale~\cite{KassRaftery} (see Table~\ref{tab:scales}).  However, these scales are not a one size fits all solution.  Of course, determining a criteria for mapping the value of a statistic and significance is arbitrary and so either of these scales could serve as the mapping between the statistic and significance. However, physics already employs a criteria to determine whether a signal is significant, the oft-quoted ``5 $\sigma$'' $p$-value. A $p$-value is the probability of observing a signal given that a null hypothesis is true.  This $p$-value is a frequentist statistic so, at first, it may seem counter intuitive to use a frequentist criteria for a Bayesian calculation.  However, it is a meaningful question to ask what is the distribution of any statistic over different realizations of the data, even the Bayesian evidence. 

\begin{table}
\centering
\begin{tabular}{|c|c|c|}
\hline
Jeffreys scale $Z_{i}/Z_{j}$ &  Kass-Rafferty scale $Z_{i}/Z_{j}$& Interpretation \\
\hline
1 to 3.2 & 1 to 3 & Not worth mentioning \\
3.2 to 10 & 3 to 20 & Positive \\
10 to 100 & 20 to 150 & Strong \\
 $>$ 100 & >150 & Very Strong \\
\hline
\end{tabular}
\caption{\label{tab:scales} The Jeffreys and Kass-Raftery scales. }
\end{table}

A third drawback is that Bayesian evidences generally depend on the subjective choice of a prior. Common choices of such a prior includes the flat prior, where the prior probability of a value of a parameter being true is uniform over some domain.  This is often a good choice since such a flat prior reflects a maximum amount of agnosticism about the value of a parameter. However, flat priors are not invariant under changes in parameterization.  e.g. flat in $P(\theta)$ is not flat in $P(\log \theta)$. Further, any uniform distribution is only well-defined if the domain is finite e.g. we might say $\Omega_{\rm m}$ is free to vary in the range $[0,1]$.  However, since the Bayesian evidence is the average of the likelihood over the prior volume, if the volume were expanded to include regions of low likelihood, then the evidence will decrease.

Previous studies have investigated the utility of Bayes factors and the Jeffreys scale for performing model selection~\cite{2008arXiv0811.2415S,2011MNRAS.413.2895J,2019PhRvD..99d3506R,2021A&A...647L...5J}. For instance,  \citet{2011MNRAS.413.2895J} treat the evidence ratio as a statistic and point out that it is noisy (see also~\cite{2021A&A...647L...5J}) and thus selecting models based solely on that number may be insufficient. Both ~\cite{2011MNRAS.413.2895J,2021A&A...647L...5J} find that the log of the evidence should have a variance that scales with the number of data points. Furthermore, \citet{2013JCAP...08..036N} calculate false positive and false negative rates when using the Jeffreys scale to select a particular class of models.  Specifically, the models they consider have predictions that are linear in their parameters. The different models considered in that paper are all nested with different numbers of parameters.  That is, model $M_2$ has one extra parameter compared to $M_1$ and for a certain value for that extra parameter, reduces to $M_1$.  The authors show that the Jeffreys scale can fail to penalize extra degrees of freedom when mock data was generated from the simpler model. They show that the frequency of this failure is influenced by the choice of prior.  The authors conclude that the Bayes factor should not be the only tool used for model selection.

Another recent, companion paper points out related short comings of using the Bayes factor. Specifically, \citet{2021arXiv211010977K} demonstrate that, if the true model is not an option when calculating Bayes factors, such a scheme will still pick some false model as being best and thus not rule it out.  They go on to show that the distribution of likelihoods from the iterative smoothing method ~\cite{shafieloo-etal06,shafieloo07} can rule out all false models~\cite{2021JCAP...03..034K}.  The current work takes inspiration from the methodology of that paper and shows how to use a models' evidence alone, not the Bayes factors, to determine if a model is a good fit to the data regardless of how well other models perform.

In contrast to questions of model selection, which are often answered by computing Bayes factors, there are questions of model validation which are answered with other statistical tools. Such ``goodness-of-fit'' tests can come in frequentist forms, such as the $\chi^2$ test, or Bayesian forms, such as the posterior predictive distribution~\cite{2021MNRAS.503.2688D}.

In this paper, we seek to develop a methodology to address these three drawbacks of using Bayes factors for model comparison. We first calculate the distribution of the Bayesian evidence, for datasets commonly used in cosmology, and show how to use this distribution to answer the question of whether a model is a good fit to the data, irrespective of other models.  Further, for these same datasets, we show how to assign a frequentist $p$-value to the Bayes factor to characterize significance when used for model comparison.

\section{Mock Data}
The approach to calculating the distribution of the Bayesian evidence is to generate a hundred mock realizations of the data and then calculate the Bayesian evidence for each one. The datasets that we use for this calculation are mock future supernova (SN) datasets as might be expected from the WFIRST telescope~\cite{green-etal12,spergel-etal15} and future baryon acoustic oscillation (BAO) datasets as might be expected from DESI~\cite{desi-etal16a,desi-etal16b}

Type Ia SN are one of the observational pillars that built the $\Lambda$CDM model and directly measure the acceleration of the Universe. Existing SN datasets have all shown broad consistency with the $\Lambda$CDM model despite concerted searches for new physics or systematics~\cite{riess-etal98,perlmutter-etal99,riess-etal07,kowalski-etal08,hicken-etal09,amanullah-etal10,suzuki-etal12,betoule-etal14,scolnic-etal18,lhuillier-etal19,koo-etal20,2021AJ....161..151K}.
These successes of SN exist because they are standardizable candles; the measurement and fitting of the light-curve of the SN allows us to infer the luminosity of the SN (up to a global calibration)~\cite{riess-etal98,perlmutter-etal99,riess-etal07,kowalski-etal08,hicken-etal09,amanullah-etal10,suzuki-etal12,betoule-etal14,scolnic-etal18}, and thus measuring the flux is equivalent to measuring a luminosity distance, or equivalently a distance moduli:
\begin{equation}
    \mu(z) = 5 \log_{10} D_L(z) + 25 = m_B(z) - M_B + \alpha x - \beta C.
\end{equation}
Here, $\mu$ is the distance modulus, $D_L$ is the luminosity distance, $m_B$ is the peak B-band flux, $M_B$ is the absolute magnitude of SN and serves as the parameter that calibrates the distance redshift relation, and $\alpha$ and $\beta$ are hyperparameters that control how the light-curve parameters ($X_1$ and $C$) influence the distance-redshift relation. In our mock dataset, we imagine the case that the SN dataset is calibrated and thus do not vary these extra parameters.  Our mock dataset includes 3000 SN in the range from $0.01<z<3.0$, and the redshifts are uniformly distributed in the $\log$ of this range.  The existing Pantheon dataset includes SN at $z=2.3$~\cite{scolnic-etal18} so it is not unreasonable that future compilations that build off of the Pantheon compilation could extend even further in redshift. The errors on the distance modulus of each SN are of the form $\sigma_{\mu} = 0.12 + 0.12 (z/3.0)^2$.
The mock datasets are generated by adding random Gaussian variables to the model predictions with standard deviation as described.  the likelihood is Gaussian when making inferences with this mock dataset.

The DESI BAO dataset measures the clustering of galaxies both along the line-of-sight and perpendicular to it and so includes constraints on both the Hubble distance ($D_H(z)$) and the angular diameter distance ($D_M(z)$) individually ~\cite{RossBAOMGS,HowlettBAOMGS,desi-etal16a,desi-etal16b,SDSSLRGBeutler,SDSSLRGAlam,SDSSLya,eBOSSLya,eBOSSCosmo}. Because the absolute distance to these galaxies is unknown, the constraints from BAO analyses are expressed as ratios of distances with respect to the sound horizon at drag epoch, $r_d$:
\begin{equation}
    \alpha_{\parallel}(z) = D_H(z) / r_d, \ \ \alpha_\perp(z) = D_M(z)/r_d.
\end{equation}
The sound horizon can be thought of as a calibration parameter, and as with the SN, we imagine a case where the calibration is known and thus our mock BAO dataset is just constraints on $D_H(z)$ and $D_M(z)$. DESI is forecasted to measure $D_H(z)$ and $D_M(z)$ in 29 individual redshift bins from $0.05<z<3.55$.  The precision of each of the measurements can vary from less than 1\% to more than 10\% (see Tables 2.3, 2.5, 2.7 in \cite{desi-etal16a}). As with the SN dataset, the mock dataset is generated by adding a random Gaussian variable to the model predictions and hence the likelihood for this dataset is Gaussian. These likelihoods that we use to generate the data are the same as the ones we use to make inferences from the data.

We generate 100 realizations of the combined datasets for each of the three models in question (see Sec.~\ref{sec:models}) for a total of 300 mock datasets.  The seed is fixed for the different models so the residuals are the same. As in, there are only 100 unique realizations of the residuals. 

\section{Models}\label{sec:models}

\begin{figure*}
    \centering
    \includegraphics[width=\textwidth]{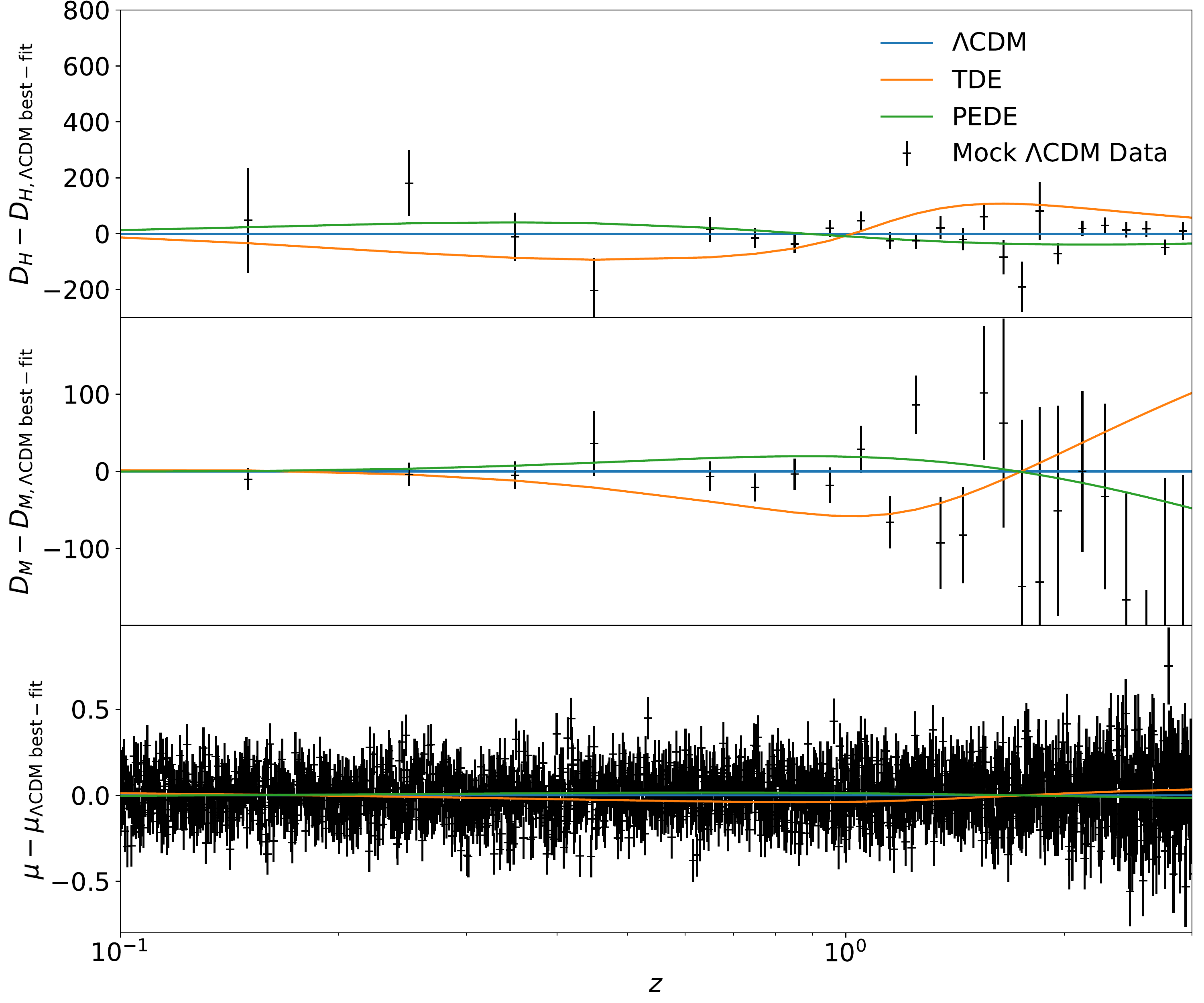}
    \caption{One example mock dataset (generated from the fiducial $\Lambda$CDM model) along with the predictions of our three considered models, $\Lambda$CDM, TDE, and PEDE, that are best fit to the data.}
    \label{fig:models}
\end{figure*}

We consider three distinct models when computing the distribution of the Bayesian evidence and Bayes factors.  Specifically we analyze the $\Lambda$CDM model, the Transitional Dark Energy (TDE)~\cite{TDE} model, and the Phenomenologically Emergent Dark Energy (PEDE)~\cite{li-shafieloo19} model.  The TDE model is parameterized thusly,
\begin{equation}
    w(z) = w_0 + (w_1 - w_0 )(-1-\tanh\left[(z-z_T)/\Delta\right]),
\end{equation}
and we choose $w_0 = -0.8$, $w_1 = -2.0$, $z_T = 1.0$, and $\Delta = 0.2$. The PEDE model has a similar form of the parameterization,
\begin{equation}
    w(z)  = -\frac{1}{3\log 10} \left({1+\tanh\left[\log_{10}\,(1+z)\right]}\right)-1.
\end{equation}
The parameters of the TDE model are fixed to put it on a level playing field with PEDE (Generalized Emergent Dark Energy or GEDE~\cite{li-shafieloo20} is an extension of PEDE that connects $\Lambda$CDM and PEDE) and with $\Lambda$CDM. Thus for each of the models, the only free parameters are $H_0$ and $\Omega_{\rm m}$. $H_0$ is allowed to vary between 60 and 80 km sec$^{-1}$ Mpc$^{-1}$ and $\Omega_{\rm m}$ is allowed to vary between 0.0 and 1.0. When generating the data, we use $H_0=70$ km sec$^{-1}$ Mpc$^{-1}$ and $\Omega_{\rm m}=0.3$ as the fiducial parameters for each of the models. These models were specifically chosen to be distinguishable given the mock datasets. This is achievable in part because, for the parameters we chose, the TDE model behaves like a quintessent ($w>-1$) dark energy in the region most constrained by the data.  On the other hand, the PEDE model exhibits a purely phantom ($w<-1$) behavior.  $\Lambda$CDM, of course, has a cosmological constant dark energy ($w=-1$). Though both of these models can be thought of as nested extensions to $\Lambda$CDM when their additional parameters are allowed to vary, these types of models are not commonly investigated alternatives to $\Lambda$CDM.  These models are further interesting since they can encode an evolution in $w(z)$ at different redshifts than $z=0$, as in the case with the Chevalier-Polarski-Linder~\cite{CP,L} (CPL) parameterization $w(z) = w_0 + w_a z/(1+z)$, which is the most common parameterization of evolving dark energy. So, in summary, we are investigating potential cases where, for example, if TDE is the true model of cosmology, calculating the Bayes factor between $\Lambda$CDM and PEDE may favor $\Lambda$CDM, but neither are the true model.  We show how to use the Bayesian evidence itself to show that both $\Lambda$CDM and PEDE are bad fits to the data in such a case.  

The evidences are calculated using \texttt{MultiNest}~\cite{2008MNRAS.384..449F,2009MNRAS.398.1601F,2019OJAp....2E..10F}, specifically the importance nested sampling algorithm, with 400 live points, a sampling efficiency of 0.3, and an evidence tolerance of 0.1.

We also investigate the case where the models in question have different degrees of freedom.  Specifically, we look at the cases where we allow the curvature to vary in our models, though the data is still generated from the case where the curvature is fixed to be 0. We call these models where the curvature ($\Omega_{\rm k}$) is allowed to vary, kTDE, kPEDE, and k$\Lambda$CDM.

\section{Results}

\begin{figure}
    \centering
    \includegraphics[width=\columnwidth]{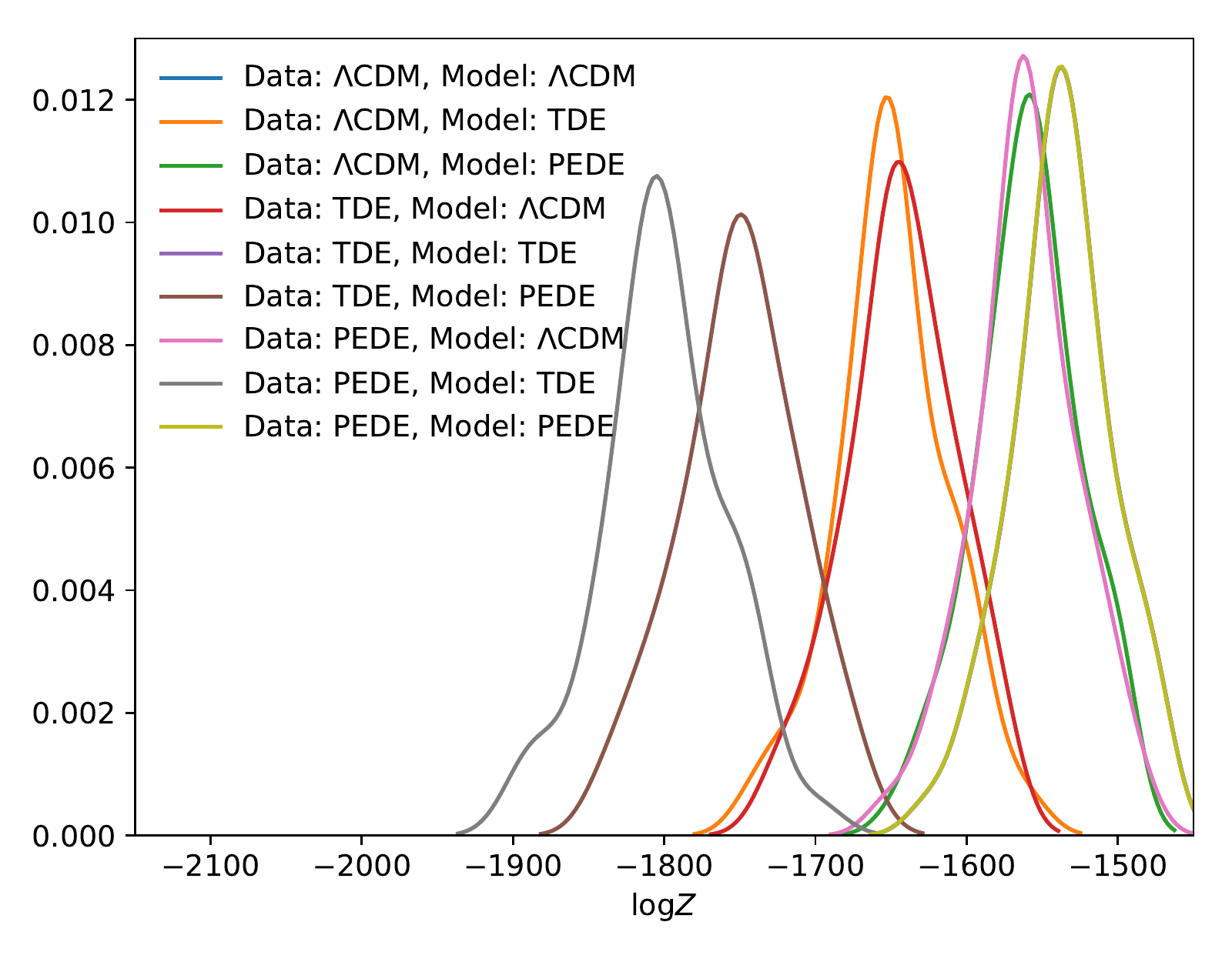}
    \caption{Distribution of the log evidences for mock datasets generated from the model labelled on the left, inferred using the model labelled on the top. Of particular note is that the distributions for the cases that the model used for the inference is the same as the one used for the data generation are all exactly the same.}
    \label{fig:dist_Z}
\end{figure}

\begin{figure}
    \centering
    \includegraphics[width=\columnwidth]{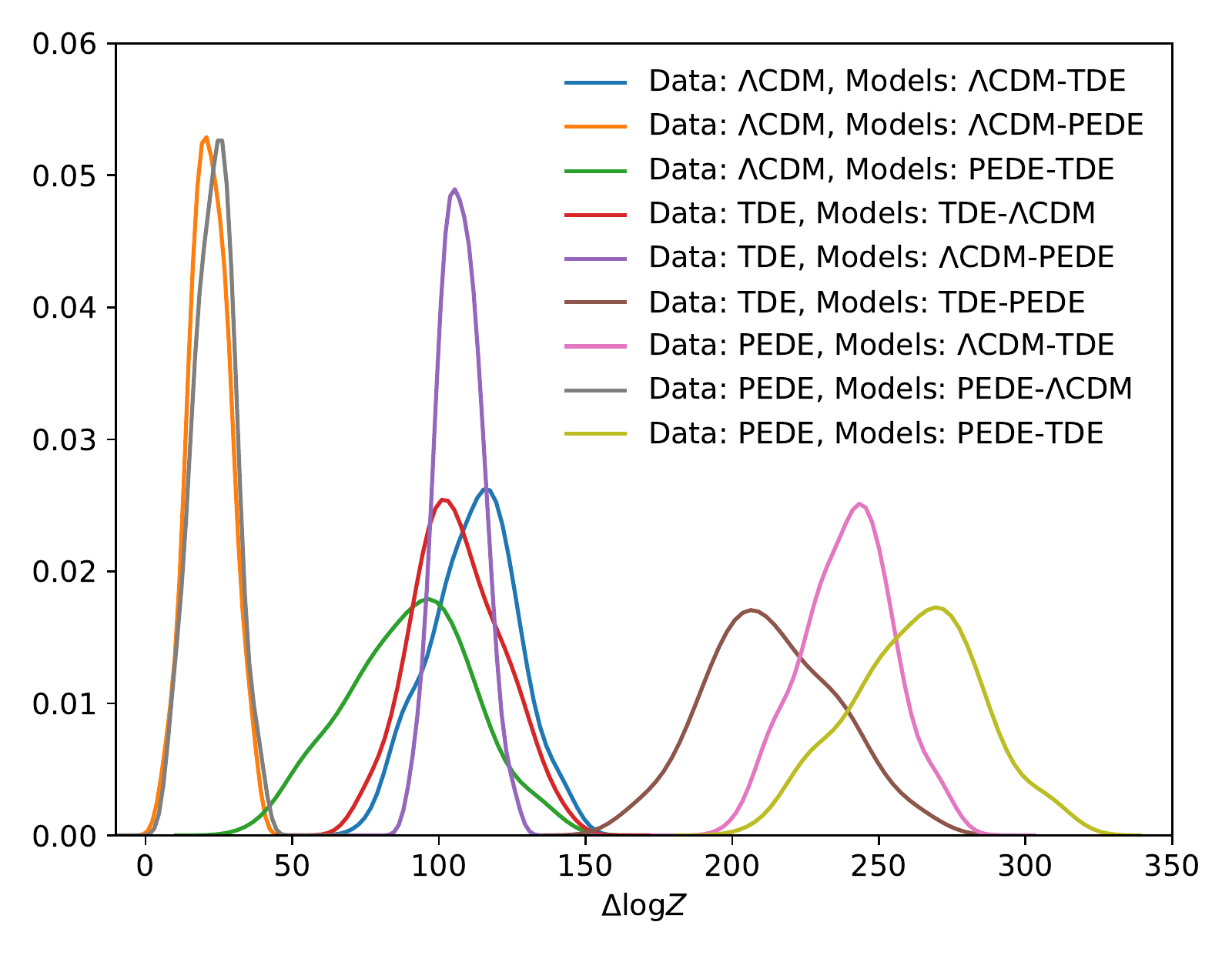}
    \caption{Distribution of the difference in the log evidence for mock datasets generated from the model labelled on the left, inferred using the models labelled on the top. The labels on the top refer the log evidence of the first model minus the log evidence of the second model.}
    \label{fig:dist_deltaZ}
\end{figure}

In Fig.~\ref{fig:dist_Z}, we see the distributions of the log evidences for nine different cases (three models were used to generate the data and three models were used for inference). 
Of particular note are the cases where the model being used to infer the evidence is the one used to generate the data.  The distributions for these cases are exactly the same and lie on top of each other in the figure.  This is expected since generating these mock datasets is essentially generating random residuals and these random residuals are the same between the different models. Additionally, the fact that the models have the same parameters that are allowed to vary is also necessary for the distributions to be identical. 
This distribution of the evidence can be used as a robust test of validation, to see if a model is a good fit to the data. To perform this test, we take the evidence calculated with real data and some model to be validated and compare it to the distribution calculated with that same model. If the evidence calculated with real data lies outside this distribution, then we can robustly say that model is ruled out or invalidated, regardless of how well other models perform.

In Fig.~\ref{fig:dist_deltaZ}, we see the distributions of the differences in the log evidences for each of the three pairwise combination of models for each of the three models used to generate the datasets. To explain the figure's labels, the difference is such that $\Delta \log Z = \log Z_{\rm L} - \log Z_{\rm R}$ so positive values represent a preference for the model on the left.  Basic expectations are borne out by this figure.  The model used to generate the data always performs better than a false model. Further, since the TDE model is a quintessent dark energy model in the regime where there is the most data, and since the PEDE model is always phantom, $\Lambda$CDM is preferred over the TDE model when the PEDE model is true, and $\Lambda$CDM is also preferred over the PEDE model when the TDE model is true.  Further, PEDE is closer to $\Lambda$CDM and thus it is preferred over TDE when $\Lambda$CDM is the true model. That $\Delta \log Z = 0$ is outside the distribution in all these cases indicates that, for this optimistic forecast for DESI BAO data and WFIRST SN data, these dark energy models should be imminently distinguishable.  One point that is worth bringing up is that the range of these distributions is noticeably different. For different model comparisons, the scatter in the $\Delta \log Z$ values will be different.  This goes to show that using a fixed scale like the Jeffreys or the Kass \& Rafferty scale cannot serve as one size fits all solutions to the interpretation of Bayes factors. Indeed, for different datasets and different models, one would need to calculate these kinds of distributions to make robust inferences about Bayes factors.

\begin{figure}
    \centering
    \includegraphics[width=\columnwidth]{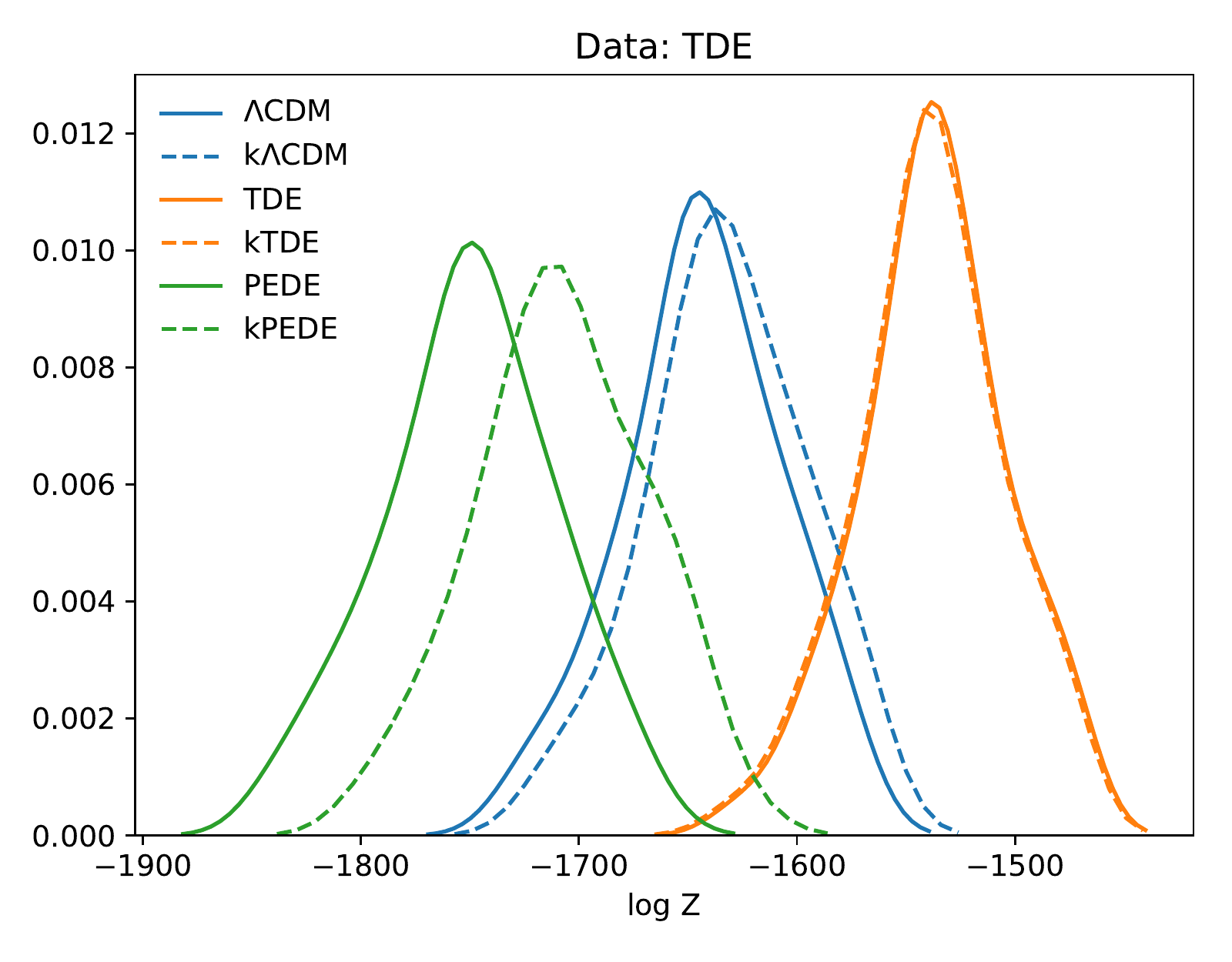}
    \caption{Distribution of the log evidences for mock realizations from the TDE model inferred using different models when curvature is and is not allowed to vary.}
    \label{fig:dist_TDE_ex}
\end{figure}

In Fig.~\ref{fig:dist_TDE_ex}, we see the distribution of evidences for the specific case of when the data is generated from the TDE model. This figure is useful to make an easy comparison to ~\cite{2021JCAP...03..034K}.  This figure also highlights one potential advantage or drawback of Bayesian methods over the distribution of likelihoods from the iterated smoothing method. The Bayesian evidence explicitly uses models and so are more interpretive and thus can say things specifically about model quantities such as curvature. However, because Bayesian methods are more interpretive, they can arrive at wrong conclusions.  As is the case in this figure, when the wrong dark energy model is used in the inference Bayesian methods can come to the wrong conclusion about the existence of curvature.

\begin{figure*}
    \centering
    \includegraphics[width=\columnwidth]{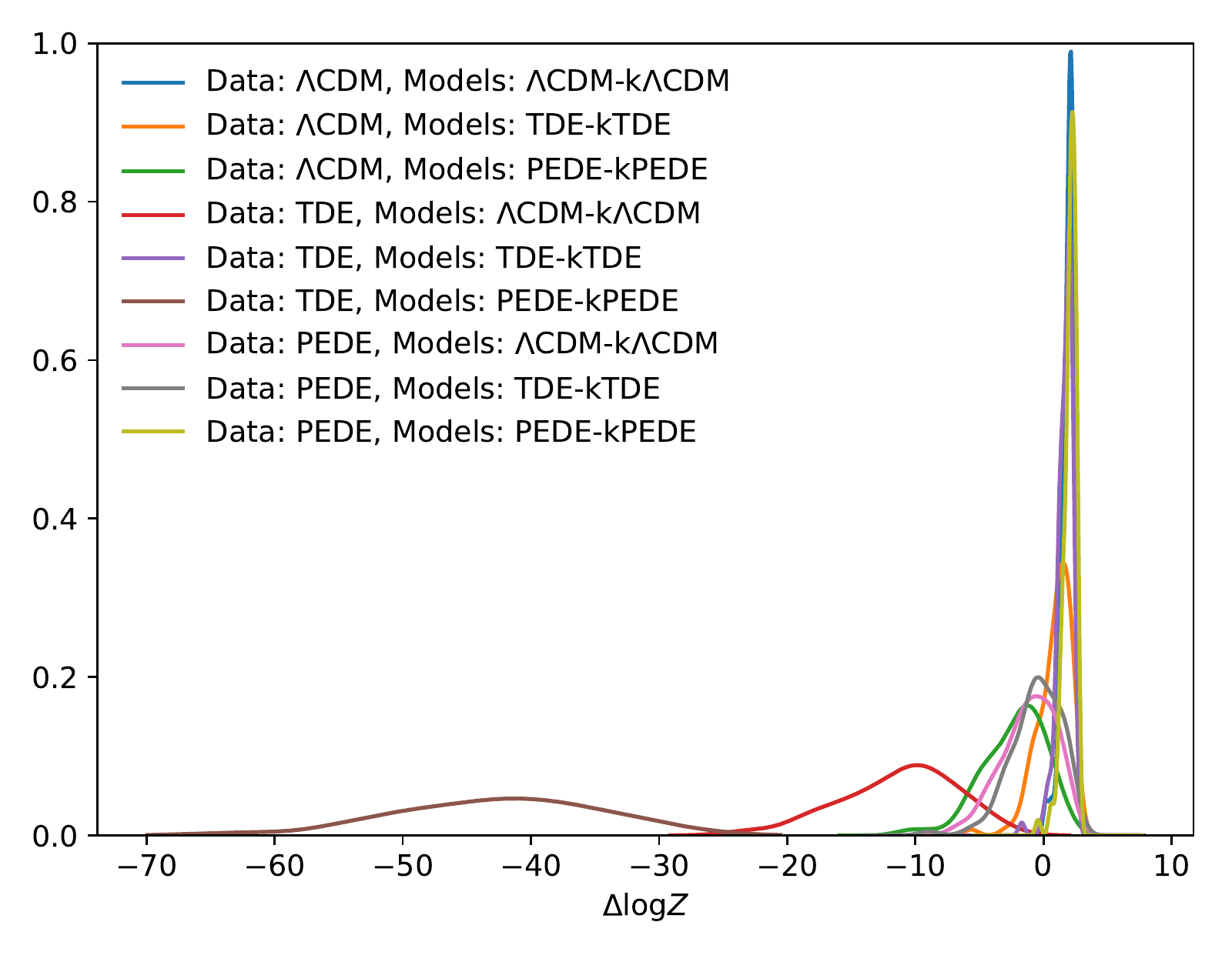}
    \includegraphics[width=\columnwidth]{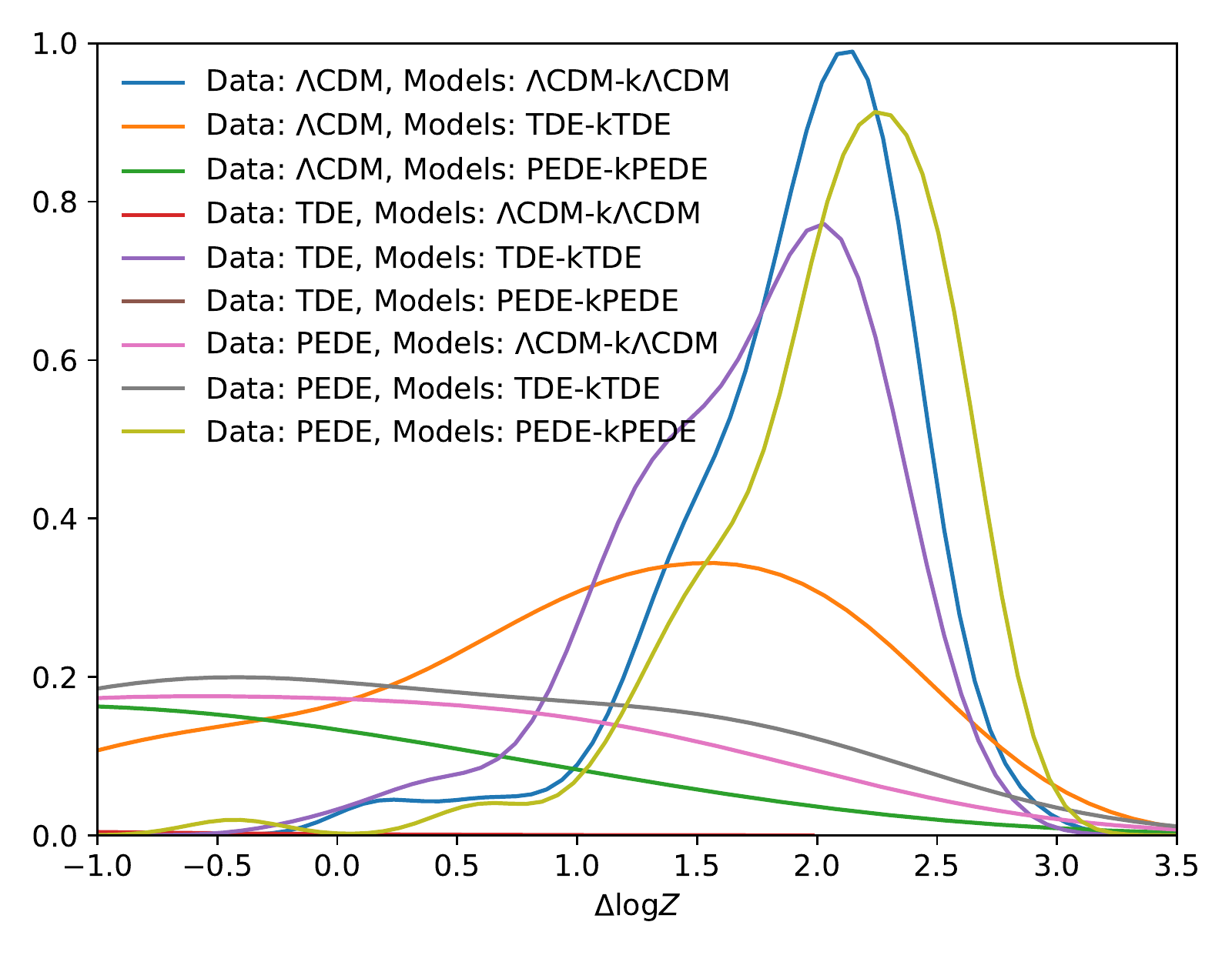}
    \caption{Distribution of the difference in the log evidence for mock datasets generated from the model labelled on the left, inferred using the models labelled on the top. The difference in the models as labelled on the top, is whether curvature is allowed to vary.  The difference in the log evidence is the difference between the flat version of the model minus the version where curvature is free to vary. The right panel is the same as the left but zoomed in on the region $-1.0<\Delta \log Z < 3.5$}
    \label{fig:dist_delta_ok}
\end{figure*}

In Fig.~\ref{fig:dist_delta_ok}, we see the distribution of the differences in the log evidences for each of the models when curvature is and is not allowed to vary. This is an example of an additional degree of freedom. In all of these cases, the data was generated from models with no amount of curvature. To explain the figure's labels, the difference is such that $\Delta \log Z = \log Z_{\rm flat} - \log Z_{\rm curved}$ and so positive values in the distribution indicate that the flat model is preferred and allowing for this extra degree of freedom is disfavored. Further, focusing on the curves where the models in question are different than the ones used to generate the data, using the wrong dark energy model can, in turn, cause wrong conclusions about whether the Universe is curved.  There can either be some amount of confusion between the mismatch in the dark energy models and the curvature (cases where the distribution spans $\Delta \log Z = 0$) or there can be a significant preference for curvature when none was included in the mock datasets, as in the cases where the data was generated with the TDE model but PEDE or $\Lambda$CDM were assumed. Further, in the cases where the model used to generate the data is the same as the model used for inference, along with its extension that includes curvature, the distributions of the evidence are roughly equivalent (they each span $\Delta \log Z \in [-1.0, 3]$). In this specific case of using a known model of dark energy and inferring whether curvature is preferred, one could set the criteria for a $>99\%$ detection of curvature at $\Delta \log Z < -1.0$. Of course, we would not a priori know that we have correctly identified the true model of dark energy. Further, this calculation does not generalize to arbitrary datasets or nested models.  For instance, if one were to include different datasets such as weak lensing or the CMB, or were to investigate different models, one would need to recalculate the significance of the  Bayes factor value.

\begin{figure}
    \centering
    \includegraphics[width=\columnwidth]{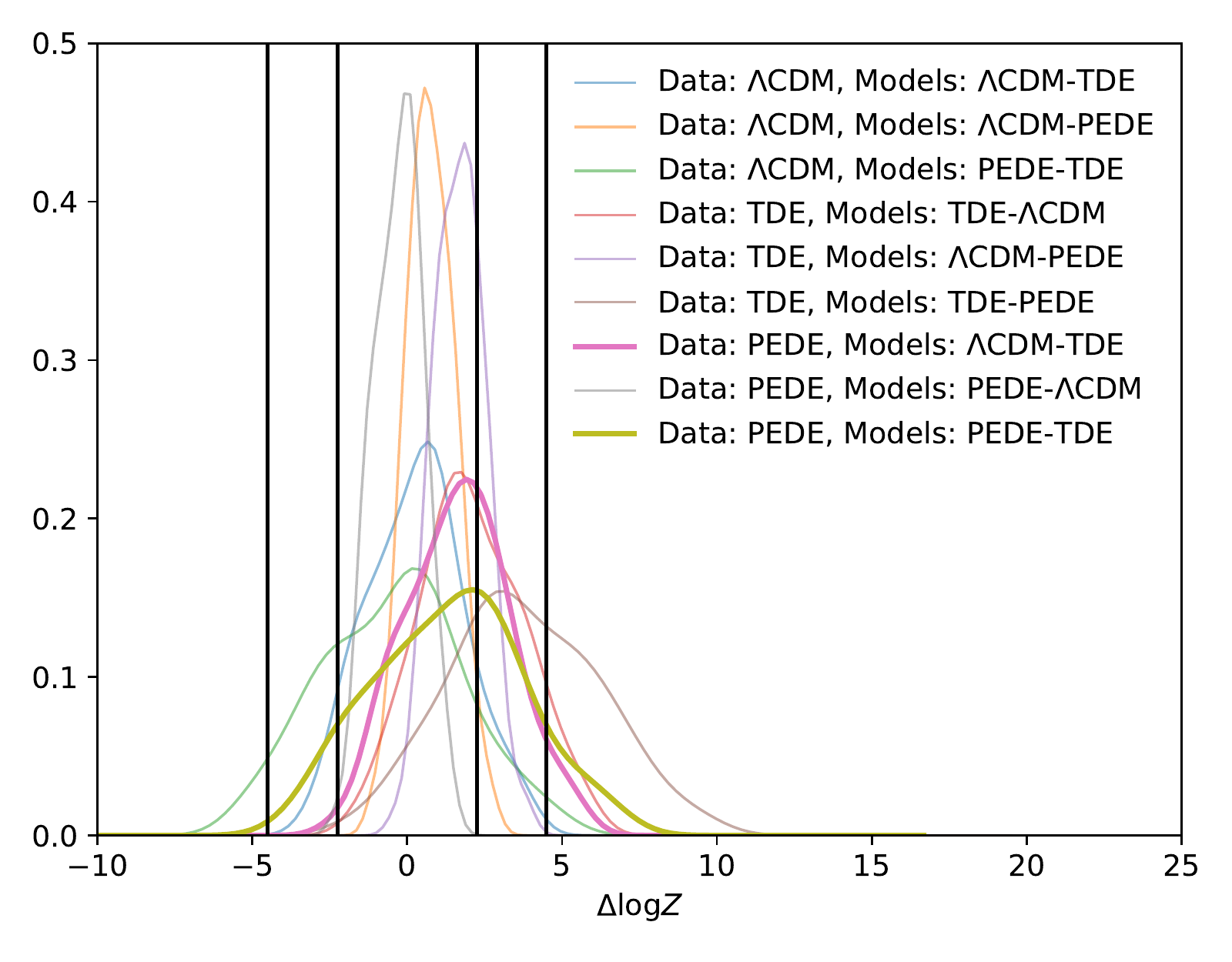}
    \caption{Same as \ref{fig:dist_deltaZ} except now the errors on the data are $10$ times larger. Since the distribution of $\Delta \log Z$ spans a wide range on either side of 0, there is a not insignificant chance of returning a "very strong" conclusion that either model is true. The pink and gold curves are highlighted as discussion will focus on them. 
    The gold curve illustrates this well where there is a non-negligible chance of returning very strong conclusions for both the correct and incorrect model, when a more indeterminate conclusion would be appropriate.
    Similarly, the case for the pink curve returns very strong or strong conclusions for two wrong models, since of course, as the true model is not included in the inference. The black vertical lines denote the ``strong'' and ``very strong'' regions of the Jeffreys scale.}
    \label{fig:BigErrs}
\end{figure}

In Fig.~\ref{fig:BigErrs}, we can more clearly see cases where the Jeffreys scale would return wrong conclusions.  The mock data used in this analysis is the same as above except now the errors are $10$ times larger. To elaborate, the data are generated with uncertainties that are 10 times larger than the normal case, and the likelihood in the analysis for this case uses the same expanded uncertainties. We will focus on the gold curve corresponding to the case where the PEDE model was used to generate the data and the two models being compared are the PEDE and TDE models. In that case in particular, 15\% of the time $\Delta \log Z$ is above $\log(100)$  and 3\% of the time $\Delta \log Z$ is below $-\log(100)$.  Thus there is a non-negligible chance that this analysis would return ``very strong'' conclusions when a more indeterminate conclusion be returned. It is additionally important to emphasize the pink curve, which corresponds to the case where the PEDE model was used to generate the data and the two models being compared are the $\Lambda$CDM and TDE models. In this case, 10\% of the time $\Delta \log Z$ is above $\log(100)$ and 3\% of the time $\Delta \log Z$ is below $-\log(10)$. Any strong conclusion is inaccurate in this case since the true model is not included in the inference. Similarly, we see for the for all of these cases the scatter in $\Delta \log Z$ is large compared the scale the different levels of the Jeffreys scale thus making it an unreliable tool for interpreting Bayes factors. The specific values for the frequencies of these confidently wrong conclusions is not of primary concern, but instead the concern arises from the fact these frequencies are non-negligible.

\section{\label{sec:Conc}Discussion and Conclusions}

The first takeaway from this paper is that the models presented here 
should be distinguishable with future WFIRST and DESI datasets.

Secondly, is that we have shown how to use the evidence itself as a criterion for model validation, to judge if a model is a good fit to the data at all, independently of any other model.  Using a distribution of the evidence for mock data generated from a model and inferred using the same model, one can see if the evidence calculated from the real data lies outside that distribution. Should the evidence computed with real data for a given model fall outside that range, then that model would be invalidated. The distribution of the evidences can be used to calculate the mapping between the Bayesian evidence and a p-value and determine a model-independent criteria to answer the question: ``Below what value would the Bayesian evidence have to be to conclude a model is not a good fit to the data?''

A third takeaway is that we have demonstrated interpretting Bayes factors using fixed scales, like the Jeffreys or Kass \& Rafferty scales, can yield incorrect conclusions. This is seen in the fact that the distributions of $\Delta \log Z$ span hundreds of values.
Even in the case where the uncertainties in the data is 10 times larger than normal, such that the models should be indistinguishable, the scatter in the Bayes factors is large compared to the range of the Jeffreys scale and will come to correct and incorrect ``very strong'' conclusions.
Thus, blindly using a fixed scale such as the Jeffreys scale, can be misleading.  To accurately assess the significance of a computed Bayes factor, one should perform these kinds of calculations, generate a sample of mock datasets and calculate the Bayes factors in those cases and then see where the Bayes factor for the real dataset lies in that distribution.

Finally, we point out that the distributions of both $\log Z$ and $\Delta \log Z$ depend greatly on the specifics of the dataset, both the size of the covariance matrix and the nature of the datasets (e.g. BAO and SN).

\section*{Acknowledgements}
We would like to thank Benjamin L'Huillier and Hanwool Koo for useful comments on the draft. This work was supported by the high performance computing cluster Seondeok at the Korea Astronomy and Space Science Institute. A.~S. would like to acknowledge the support by National Research Foundation of Korea NRF-2021M3F7A1082053 and Korea Institute for Advanced Study (KIAS) grant funded by the government of Korea.

\section*{Data Availability}
The data used in this work are mocks generated by the authors, as described in the text. They are available upon request.

\bibliographystyle{mnras}
\bibliography{BayesDist}

\appendix

\section{Scalings with Size and Types of Datasets}

\begin{figure*}
    \centering
    \includegraphics[width=\columnwidth]{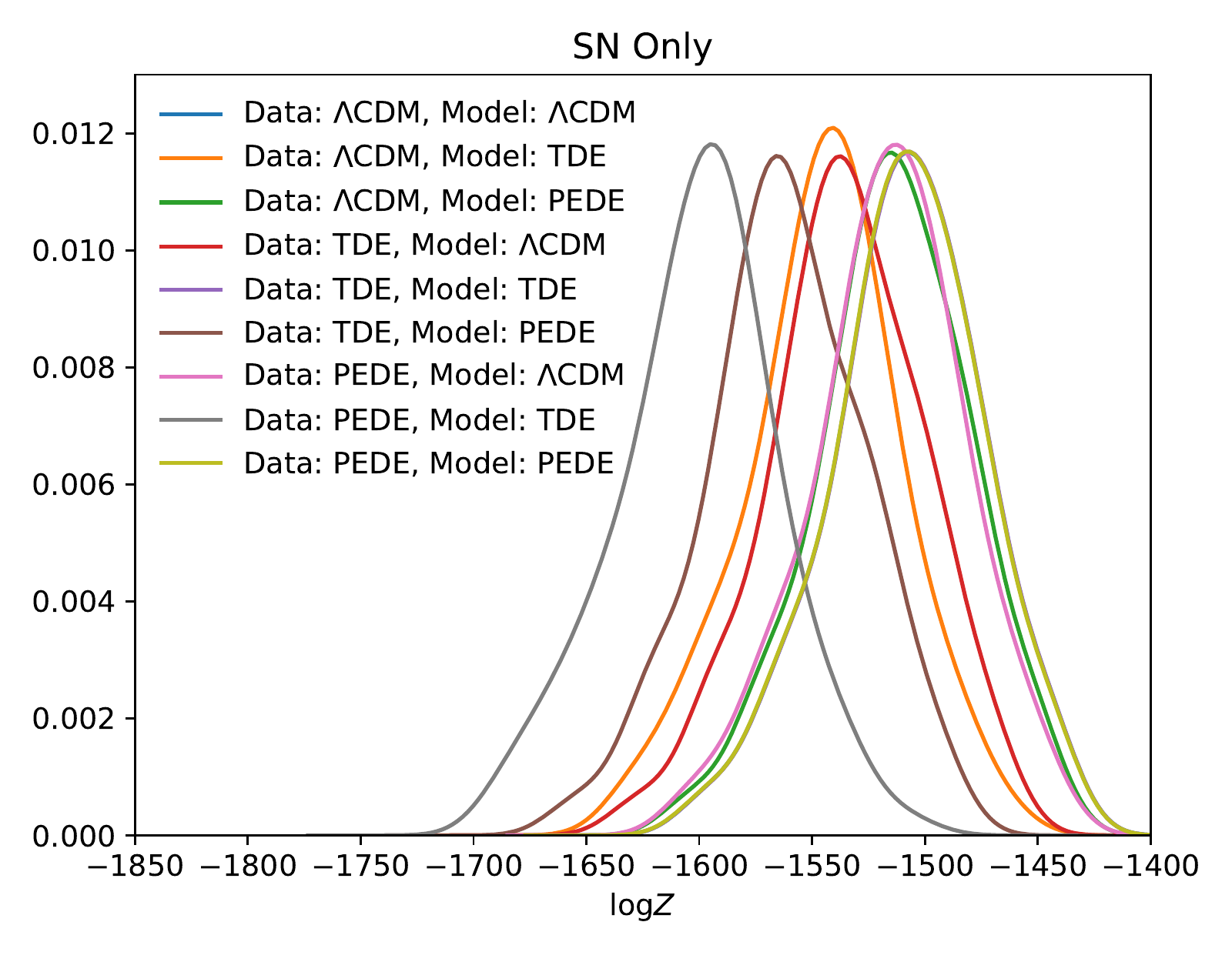}
    \includegraphics[width=\columnwidth]{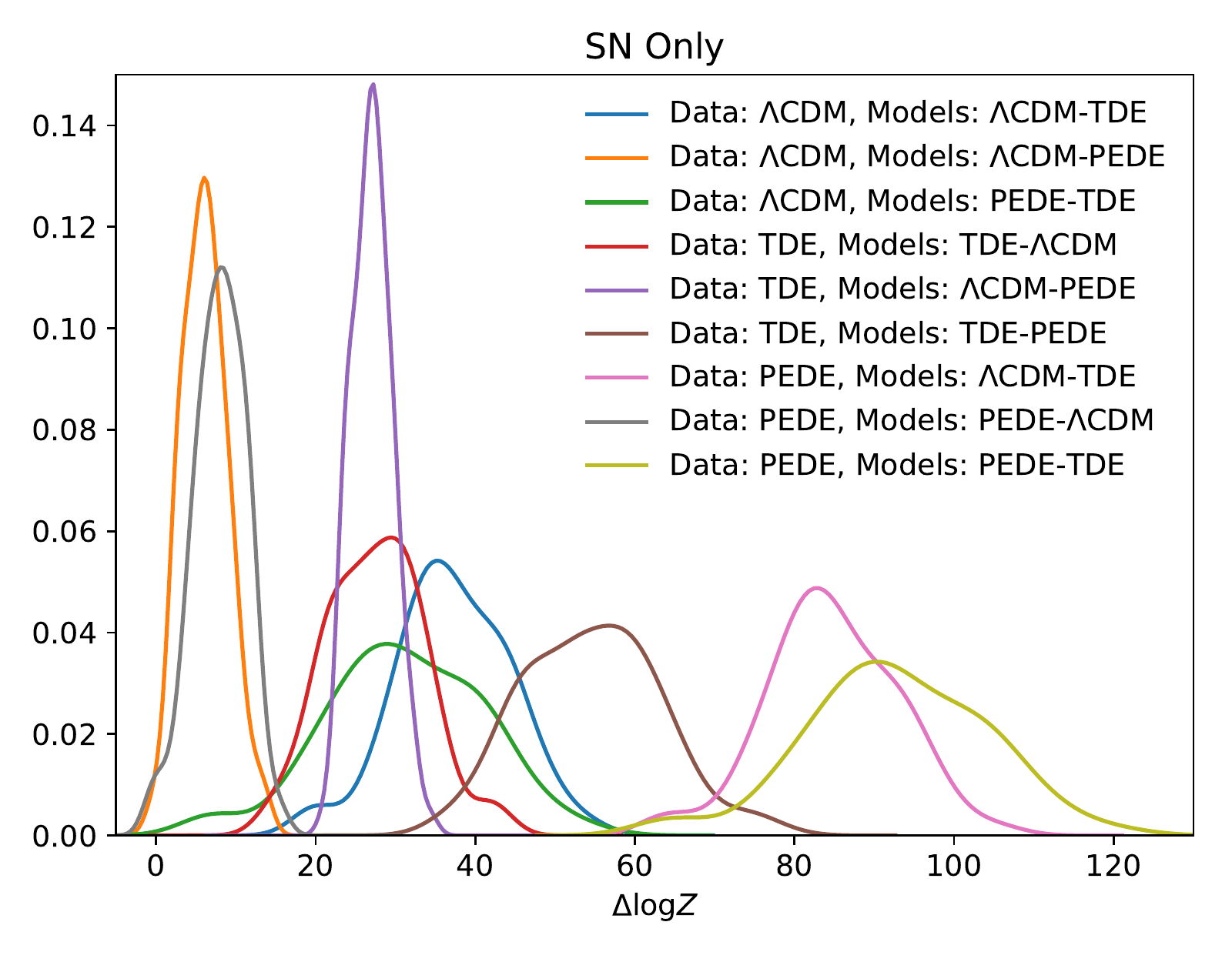}\\
    \includegraphics[width=\columnwidth]{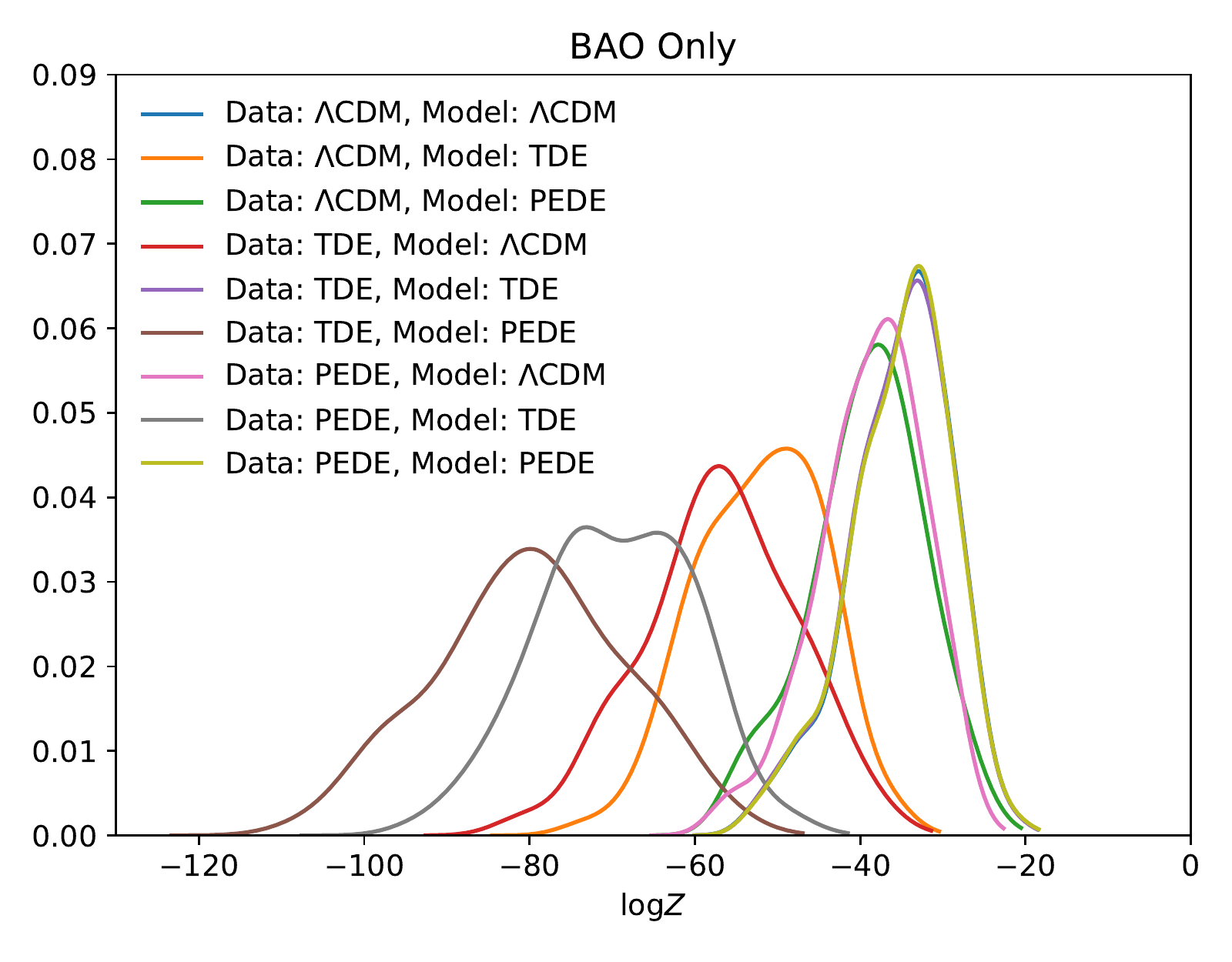}
    \includegraphics[width=\columnwidth]{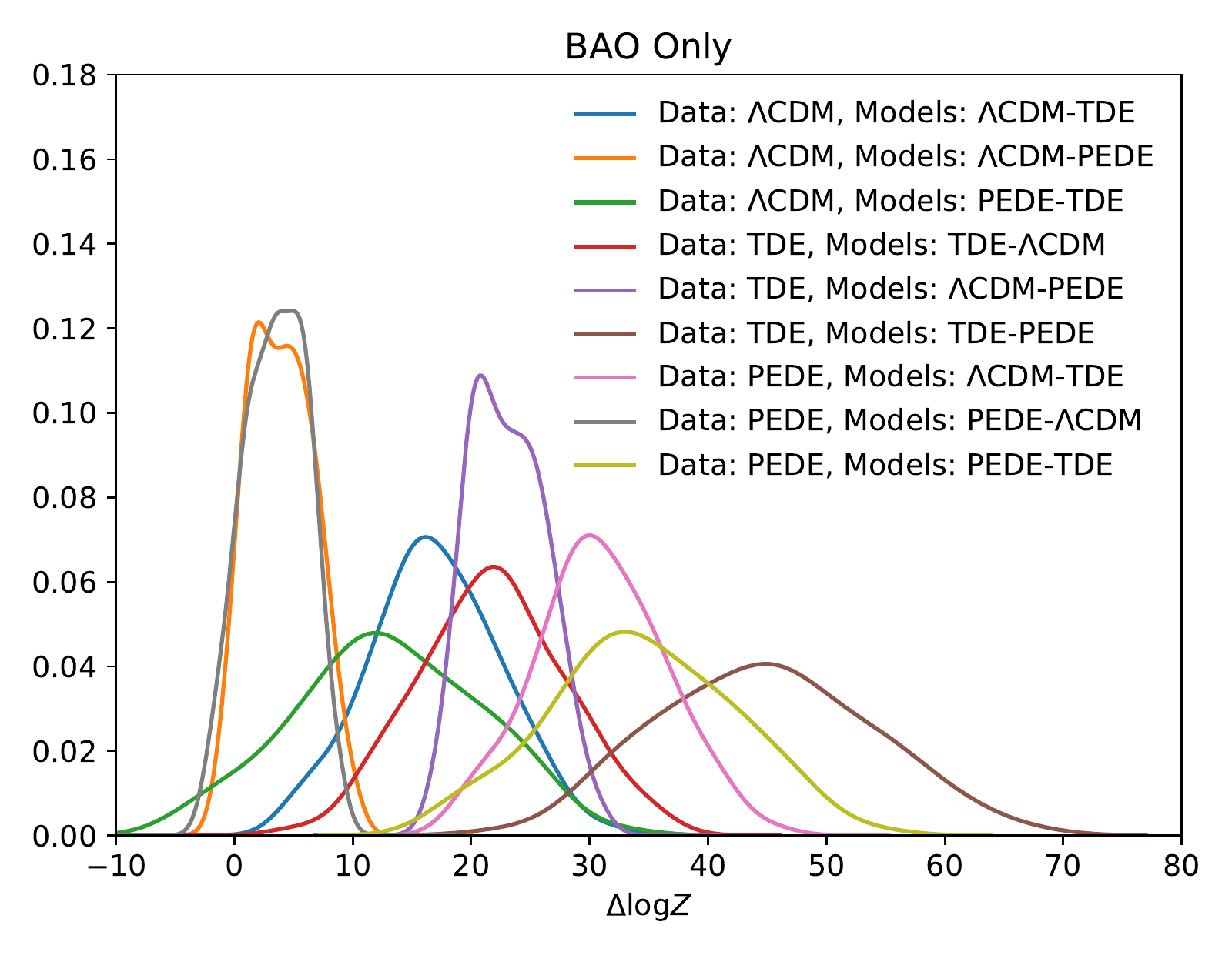}\\
    \includegraphics[width=\columnwidth]{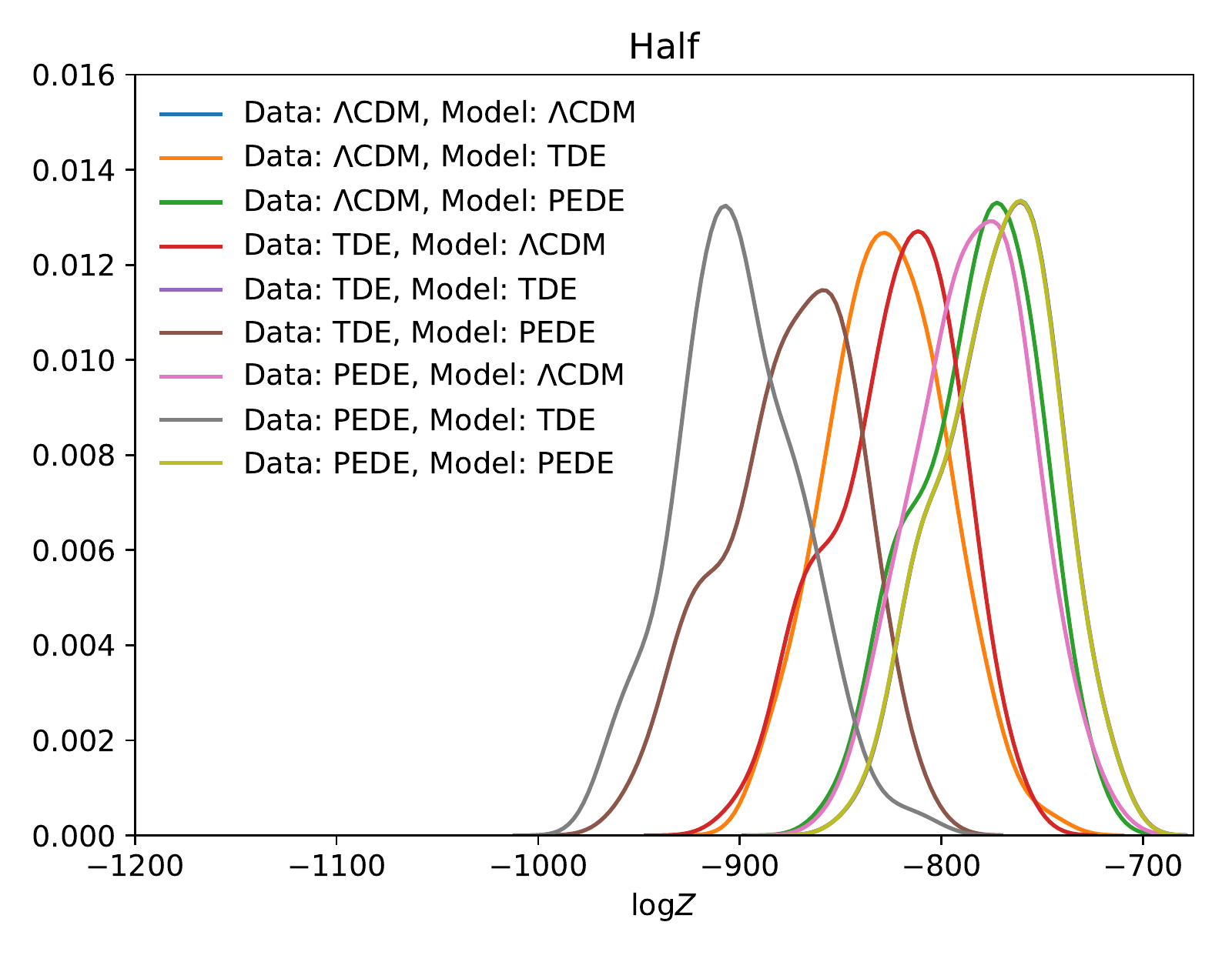}
    \includegraphics[width=\columnwidth]{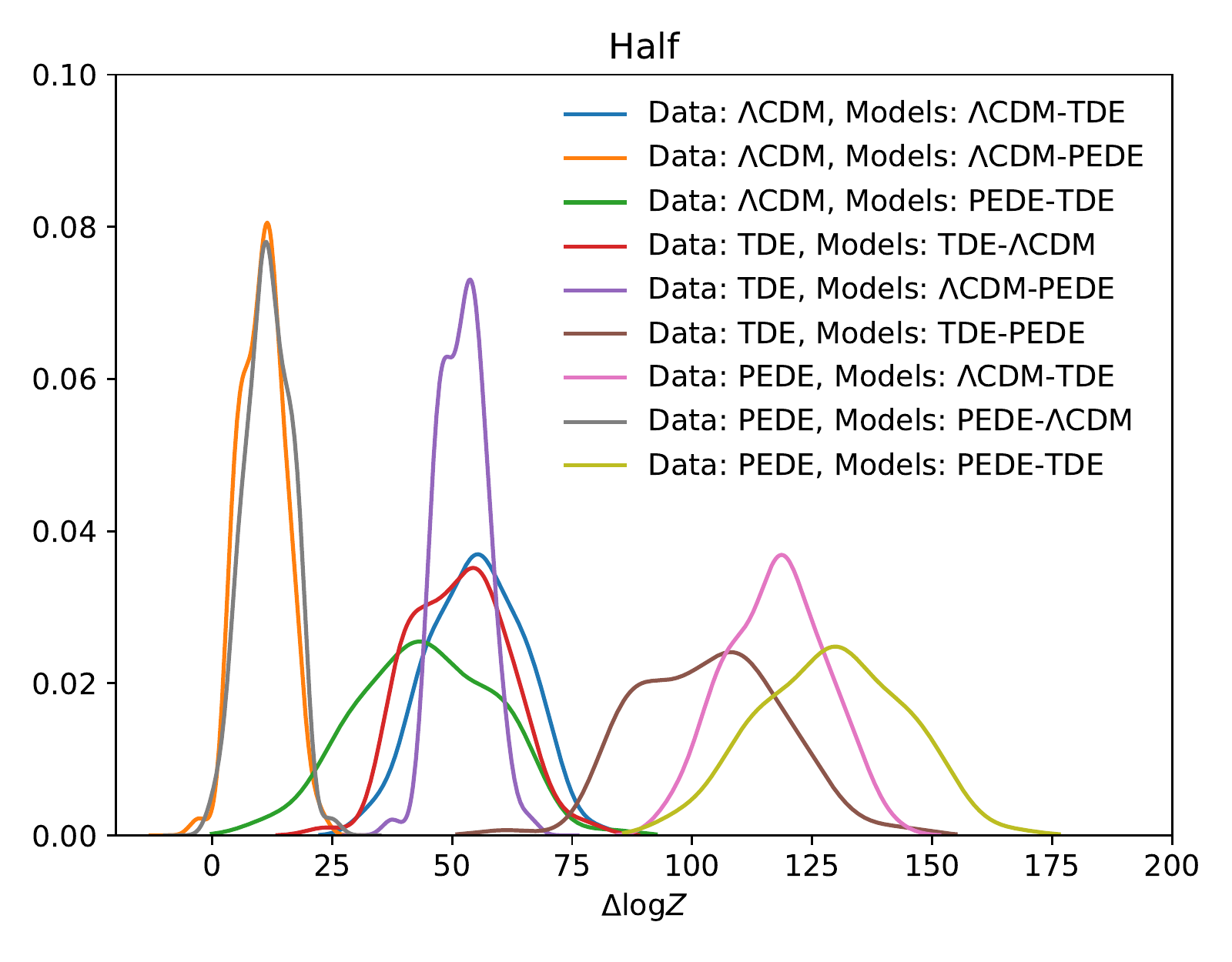}
    \caption{Distributions of $\log Z$ (left) and $\Delta \log Z$ (right) for different partitions of our full dataset.  On the top is the distributions of evidences for only the SN dataset, on the middle is the distribution of evidences for only the BAO dataset, and on the bottom is the distribution of evidences for half of the full dataset such that every other data point is used. The difference in scale corresponds to the difference in the number of data points in the different partitions.}
    \label{fig:scaling}
\end{figure*}

In this appendix, we will seek to understand the dependence of the mode and variance of the distributions of $\log Z$ on the size and kind of the dataset.  We will focus discussion primarily on the distributions where the models used to generate the data and make inferences from the data are the same since they should be the most generalizable case.

In Fig.~\ref{fig:scaling} we can begin to see the dependence of the kind and size of datasets on the distribution of evidences. We lay out three cases where we have calculated evidences for subsets of the full dataset, namely a ``SN only'' case, a ``BAO only'' case, and a case where we use every other data point for both kinds of datasets, which we will refer to as the ``half'' case. To elaborate about the ``half'' case, we include both SN and BAO information but half of the total data points of each. For the SN part of the ``half'' case, we include the distance moduli information of the lowest redshift SN, but skip the next lowest, include the third lowest, and so on.  The BAO part is similar where we include the measurement of $H(z)$ and $D_M(z)$ for the lowest redshift bin, skip the next bin, include the third bin, and so on.  
This case is the simplest comparison with the full case,
since both the BAO and SN datasets are included and since both span similar redshift ranges. In other words, this case should be most similar to a case with half the information and the transformation $N$ $\rightarrow$ $N/2$.
The point the ``SN only'' and ``BAO only'' cases is to demonstrate that the distributions of evidences become more distinguishable than would be expected from just increasing the number of data points.  This would have to arise from the fact that degeneracies in model parameters are being broken by the different kinds of datasets.

For the full dataset, the mode of the in-kind distributions are -1537 and they have a variance of 1107.  For the ``half'' dataset, the mode of the in-kind distributions are -760 and they have a variance of 758.  For the ``BAO only'' and ``SN only cases we see the mode of the in-kind distributions are -33 and -1506, respectively and their variances are 36 and 1109, respectively.
From this, we see that the difference in the modes of the various cases scales as half the number of data points $N/2$. This result is similar to the familiar statistics of measurements of multiple Gaussian random variables, with the familiar expectations that the $\chi^2$ value of the true value scales as $N/2$. 

Further the difference between the left panels of Fig.~\ref{fig:scaling} and Fig.~\ref{fig:dist_Z}, in particular that the distributions are more distinguishable in Fig.~\ref{fig:dist_Z}, arises from the fact that, more than just adding more data, the fact the BAO also constrains $H(z)$ allows it to break degeneracies that is essential for making these distributions distinguishable.  In particular, the ``SN only'' panel of Fig.~\ref{fig:scaling} has essentially the same number of data points as the full case ($3000$ vs $3058$) but the full case is far less distinguishable.  Indeed, the ``SN only'' has a similar distinguishability as the ``half'' case (see Table~\ref{tab:distinguish}). Counter-intuitively, the ``BAO only'' case is more distinguishable than the ``half'' case and is similar to the full case. Simply adding more of the same kind of data points increases the variance of these distributions more than they separate the modes of the distributions.

\begin{table}
\centering
\begin{tabular}{|c|c|c|}
\hline
Case &  Number of data points& Mode of $\log Z$ distribution \\
\hline
Full & 3058 & -1537 \\
Half & 1530 & -760 \\
SN Only & 3000 & -1506 \\
BAO Only & 58 & -33 \\
\hline
\end{tabular}
\caption{\label{tab:modes} The number of data points for each of the data cases, along with the mode of the in-kind distributions. }
\end{table}

\begin{table}
\centering
\begin{tabular}{|c|c|c|c|c|}
\hline
Data: Model: & Full & Half & SN Only & BAO Only \\
\hline
D:$\Lambda$CDM M:TDE  & 91\%  & 57\% & 19\% & 69\%  \\
D:$\Lambda$CDM M:PEDE & 11\%  & 14\% & 4\%  & 11\%  \\
D:TDE M:$\Lambda$CDM  & 85\%  & 48\% & 11\% & 83\%  \\
D:TDE M:PEDE          & 100\% & 98\% & 34\% & 100\% \\
D:PEDE M:$\Lambda$CDM & 12\%  & 11\% & 5\%  & 8\%   \\
D:PEDE M:TDE          & 100\% & 99\% & 80\% & 100\% \\
\hline
\end{tabular}
\caption{\label{tab:distinguish} The frequency that the not in-kind distributions are beyond the 95\% level of the in-kind distributions. }
\end{table}

For completeness, we include the distributions of $\Delta \log Z$ but we will limit discussion of their scalings because these distributions will naturally depend on not just the number of data points but the differences in what the models predict and thus are not generalizable. We will point out that the variance of these distributions is large, as in the full case, and, for some cases, will occasionally come to strong incorrect conclusions, as in the case with expanded errors.

\bsp	
\label{lastpage}
\end{document}